# Classification of Skin Disease Using Transfer Learning in Convolutional Neural Networks

Jessica S. Velasco[1], Jomer V. Catipon[2], Edmund G. Monilar[3], Villamor M. Amon[4], Glenn C. Virrey[5], Lean Karlo S. Tolentino[6]

[1,2,3,4,5,6]Department of Electronics Engineering, College of Engineering, Technological University of the Philippines, Manila, Philippines
[5]Department of Electronics Engineering, University of Santo Tomas, Manila, Philippines
[6]Center for Artificial Intelligence and Nanoelectronics, Integrated Research and Training Center, Technological University of the Philippines, Manila, Philippines

*Abstract*— Automatic classification of skin disease plays an important role in healthcare especially in dermatology. Dermatologists can determine different skin diseases with the help of an android device and with the use of Artificial Intelligence. Deep learning requires a lot of time to train due to the number of sequential layers and input data involved. Powerful computer involving a Graphic Processing Unit is an ideal approach to the training process due to its parallel processing capability. This study gathered images of 7 types of skin disease prevalent in the Philippines for a skin disease classification system. There are 3400 images composed of different skin diseases like chicken pox, acne, eczema, Pityriasis rosea, psoriasis, Tinea corporis and vitiligo that was used for training and testing of different convolutional network models. This study used transfer learning to skin disease classification using pre-trained weights from different convolutional neural network models such as VGG16, VGG19, MobileNet, ResNet50, InceptionV3, Inception-ResNetV2, Xception, DenseNet121, DenseNet169, DenseNet201 and NASNet mobile. The MobileNet model achieved the highest accuracy, 94.1% and the VGG16 model achieved the lowest accuracy, 44.1%.

## I. INTRODUCTION

Skin diseases are defined as conditions that typically develop inside the body or on the skin and manifest outside. There are 3000 types known skin disease [1]. Some conditions are uncommon while some occurs commonly. Generally, this condition brings itch, pain, and sleep deprivation. Other effects of skin diseases include emotional and social impact due to its detectable visual sensation. However, dermatologists assures that majority of skin diseases can be controlled with proper medication and properly diagnosed.

An implementation of an accurate and precise automated skin disease detection application that can be used by dermatologists can help in reducing their job.

Big Data refers to gathering and processing dataset sets to the level where its size and complexity transcends the capacity of conventional data processing applications. It is distinguished by 5Vs: (1) huge volume of data, (2) wide variety of data types, (3) velocity of data processing, (4) variability of data, and (5) value of data [2]. Some of the repositories available online includes molecular, clinical, and epidemiology data. This provides a vast possibility for research opportunities for different scientific advancements [3]. By combining the use of big data, image recognition technology, and the field of dermatology, patients, dermatologists, and the research community might reap a great benefit. This is due to many skin diseases that be diagnosed by medical professionals by inspecting it with naked eye. The different visual feature of each condition makes them easy to diagnose with the use of artificial intelligence and deep learning technologies. Moreover, skin diseases that are common in the Philippines can be easily identified with the of image recognition technologies. These skin diseases include chicken pox, acne, eczema, pityriasis rosea, psoriasis, tinea corporis, and vitiligo.

Previously, MobileNet was only used as the model in skin disease classification [4]. In this paper, additional learning models are implemented such has VGG16, VGG19, Xception, ResNet50, InceptionV3, InceptionResNetV2, DenseNet121, DenseNet169, DenseNet201, and NASNet Mobile. They were used to classify skin diseases with the use of images gathered from websites that are professional and open to public use like photo atlas of dermatology. They were tested if they will outperform the previously implemented MobileNet.

## II. Conceptual Literature

### A. Transfer Learning

Human learners have the ability to naturally transfer their knowledge between one task to another. In other words, when faced with new challenges, people can recognize and use the pertinent information from past experiences. The ease of learning a new task depends on how resembles our previous knowledge. Contrarily, typical machine learning algorithms focus on small tasks. Transfer learning aims to change this b y creating strategies to use knowledge acquired in one or more source activities and apply it to enhance learning in a related target activity. To make machine learning as effective as human learning, knowledge transfer techniques are being pushed to advance [5].

### B. Keras Platform

A Fully Convolutional Network (FCN) was implemented, designed and developed using Keras, Python, and Theano in the research "Fully convolutional networks for segmenting pictures from an embedded camera" [6]. The FCN is used in this research to perform basic computer vision operations on images from a robot-mounted small stereo imaging sensor.

The network's design was prototyped using the Keras library, which accelerated the search for a network with high accuracy and minimal computing resource usage. The dataset of images is modified to fit the stereo camera imaging acquisition presets for the robot. It was also used for the training and validation of the proposed network.

### C. Inception V3

The Inception-v3 model of the Tensor Flow platform was used by the researchers in the study "Inception-v3 for flower classification" [7] to categorize flowers. The flower category dataset was retrained using transfer learning technology, which can significantly increase flower classification accuracy. In comparison to previous methods, the model's classification accuracy was 95% for the Oxford-17 flower dataset and 94% for the Oxford-102 flower dataset.

### D. MobileNet

Researchers utilized a Convolutional Neural Network model called MobileNet in the study "Driver distraction detection using single convolutional neural network" [8] to identify driver distraction.

The findings for MobileNet accuracy are seen to be higher compared to the Inception ResNet . Moreover, system results vary widely depending on how quickly CPU/GPU processing time is.

### E. Inception-ResNet-V2

The researchers introduced a brand-new family of modules called the PolyInception in their paper "PolyNet: A Pursuit of Structural Diversity in Very Deep Networks" [9]. It can replace various network components in a composition or isolated form with flexibility. Architectural efficiency can be used to choose PolyInception modules to increase expressive capability while maintaining a similar computational cost.

Based on the Inception-ResNet-v2 has the highest documented single model accuracy on ImageNet. Inception blocks are utilized to capture the residuals in the most recent version of the residual structure, which combines the two. The building blocks of GoogleNet are inception blocks. Their structures have undergone multiple iterations of optimization and refinement.

### F. VGG-16

Different CNNs have been introduced with various architectural designs. With lower convolutional size and strides, the VGG-16 consists of 16 layers (13 convolutional layers and 3 fully linked layers). 4096 channels are present in the first two fully connected layers, while 1000 channels are present in the third layer. With the exception of sampling the inputs from the cropped multi-scale training images, VGG 16 uses a nearly identical training process to AlexNet. Using a convolutional neural network, the marine industry can recognize visual objects [10].

### G. VGG-19

The researchers suggested a method to help the blind by delivering contextual information of the surroundings using 360° view cameras mixed with deep learning in the study "360° view camera based visual assistive technology for contextual scene information" [11]. The feed gives the user contextual information in the form of audio. That is accomplished by utilizing CNN transfer learning properties with the pre-trained VGG-19 network to classify data using convolutional neural networks (CNN).

VGG-19 convolutional neural network is a 19-layers network. It is composed of convolutional layers, Maxpooling, fully connected layers, and an output Softmax layer.

The network's design is sequential, which means that layers are arranged in a stack. The online collected photos dataset was used for training and testing as a proof of concept for the VGG-19 in combination with a linear classifier to handle the classification challenge for this paper.

*H. DenseNet*

DenseNet, developed by Gao Huang and colleagues, connects each layer to every other layer in a feed-forward method. Unlike standard L-layer convolutional networks, which contain L connections—one between each layer and its succeeding layer—our network has L(L+1) 2 direct connections. The feature-maps of all preceding layers are utilized as inputs for each layer, and its own feature-maps are used as inputs for all following layers. DenseNets have several appealing advantages: they solve the vanishing-gradient problem, improve feature propagation, increase feature reuse, and reduce the number of parameters significantly. Our suggested architecture is evaluated using four highly competitive object recognition benchmark tasks (CIFAR-10, CIFAR-100, SVHN, and ImageNet) [12].

### III. METHODOLOGY

*A. Dataset*

The photos required for the project's development are sourced from the www.dermweb.com photo atlas, notably www.dermnetnz.org, as well as many clinical dermatological photo atlas publications. Acne, Varicella (chickenpox), eczema, Pityriasis rosea, psoriasis, vitiligo, and Tinea corporis are examples of skin diseases depicted in Figure 1. The datasets were compiled using a combination of publicly accessible dermatological repositories, dermatology color picture atlases, and photographs acquired by hand. Dermatologists have approved them as the categorization of skin disorders.

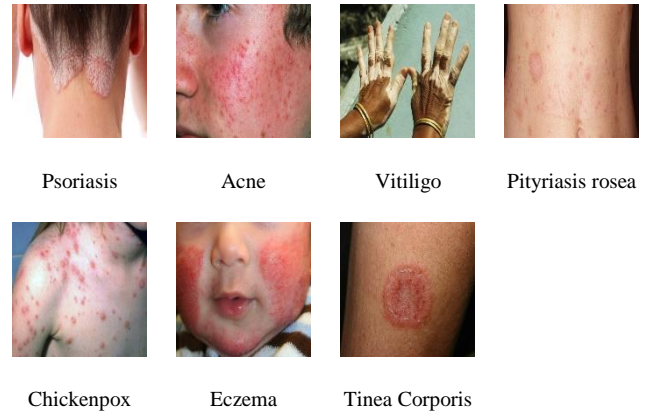

Psoriasis    Acne    Vitiligo    Pityriasis rosea

Chickenpox    Eczema    Tinea Corporis

**Figure 1. Sample Images of Dataset**

The dataset comes from a combination of open-access dermatological website, color atlas of dermatology and taken manually. The dataset composed of 7 categories of skin diseases and each image is in .jpeg extension. There is a total of 3,406 images.

*B. Experiment*

The system will be built on the Keras platform and will use Tensorflow as its backend. The Pycharm IDE will be used to develop the app. The method can detect skin problems such as acne, eczema, psoriasis, vitiligo, Tinea corporis, chicken pox, and Pityriasis rosea. This is accomplished through the use of convolutional neural network transfer learning models such as the VGG 16, VGG 19, Inception, Xception, ResNet50, DenseNet, and Mobilenet.

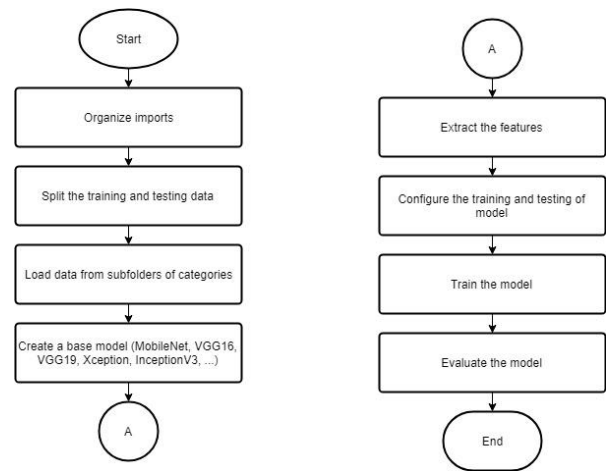

**Figure 2. Program flowchart of the Python Code**

Referring to Figure 2, imports such as Numpy, Keras, Scikit-Learn, and Matplotlib are organized first by the application.

The dataset should then be configured into several directories to separate the training and testing data (training, testing and validation). The third step is to load photographs of skin conditions from category subfolders. Making a foundation model of various pretrained convolutional neural networks is the next step. Next, the data is preprocessed to get the features. To handle this automatically, Keras includes tools. The model's testing and training configuration comes next. The model is trained using the Adam optimizer. In order to determine which architecture is optimal for classifying skin diseases, various architectures will be assessed and compared based on model accuracy, confusion matrix, loading time, and weight size after training.

Using pretrained convolutional networks, size of the input image differs for each model. The input image is equal to the size of the image (width and height) and the number of channels. Table I shows the fixed size of the input image for each model.

**TABLE I**
**INPUT FOR EACH MODEL**

| Model | Input Image |
|---|---|
| MobileNet | 224x224x3 |
| VGG16 | 224x224x3 |
| VGG19 | 224x224x3 |
| Xception | 299x299x3 |
| ResNet50 | 224x224x3 |
| InceptionV3 | 299x299x3 |
| InceptionResNetV2 | 299x299x3 |
| DenseNet121 | 224x224x3 |
| DenseNet169 | 224x224x3 |
| DenseNet201 | 224x224x3 |
| NASNet Mobile | 224x224x3 |

## IV. RESULTS

The following criteria were looked at to compare and validate each pre-trained convolutional neural network's performance in classifying skin diseases: the confusion matrix, loading speed, accuracy, and weight size.

*A. Confusion Matrices*

The confusion matrices of several models over the seven types of skin diseases are displayed in Figures 3-12. The row denotes a projected class, while the column denotes the actual class [13]. It is also known as a matching matrix. This demonstrates the commonality in misclassification across several convolutional neural networks.

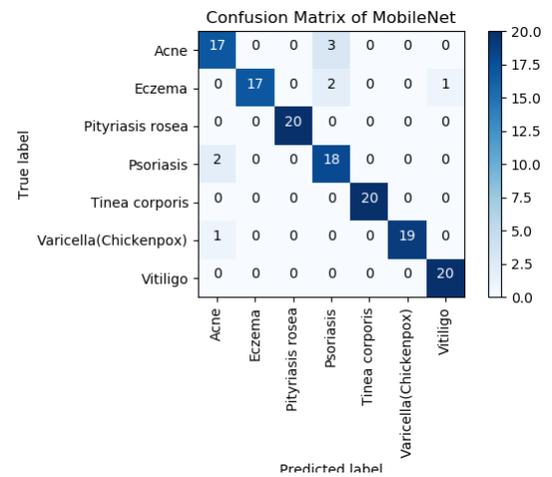

**Figure 3. MobileNet Confusion Matrix**

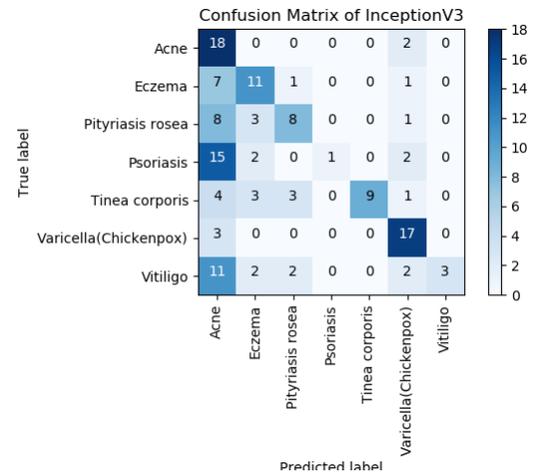

**Figure 4. InceptionV3 Confusion Matrix**

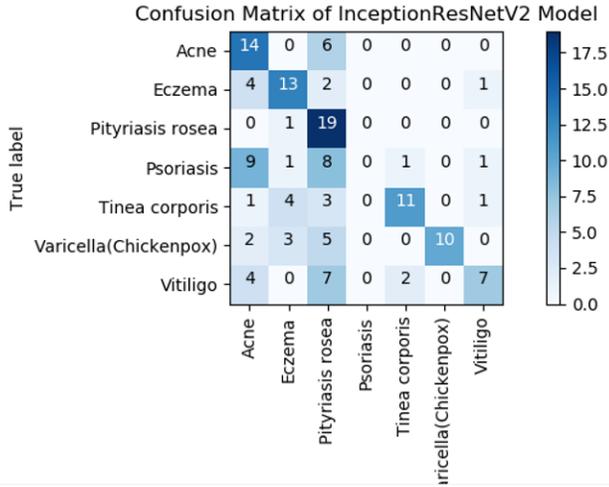

**Figure 5. InceptionResNetV2 Confusion Matrix**

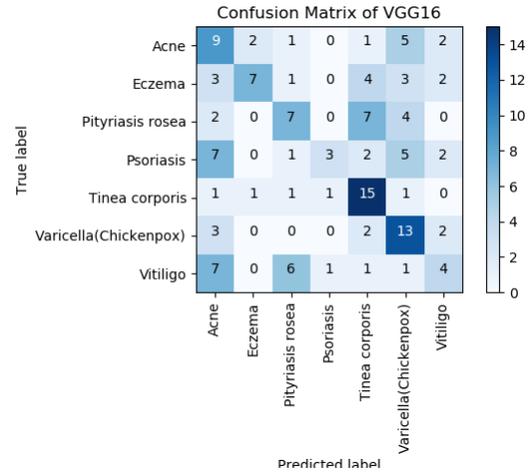

**Figure 7. VGG16 Confusion Matrix**

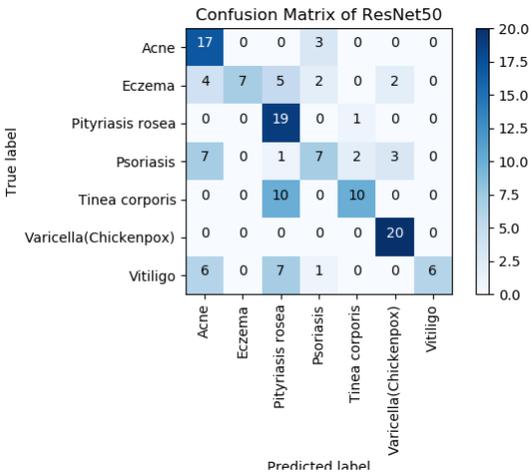

**Figure 6. ResNet50 Confusion Matrix**

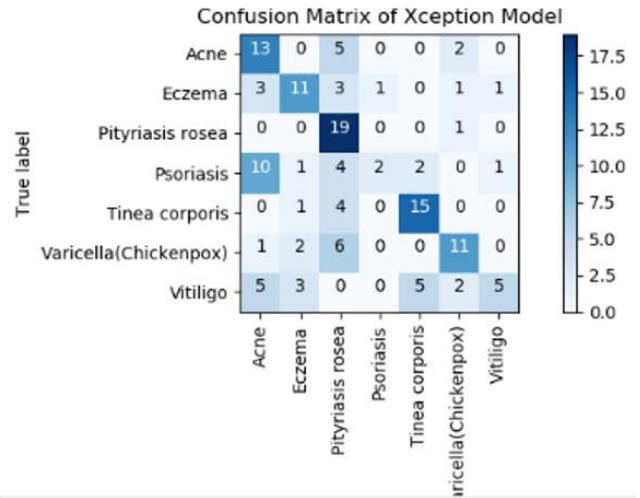

**Figure 8. Xception Confusion Matrix**

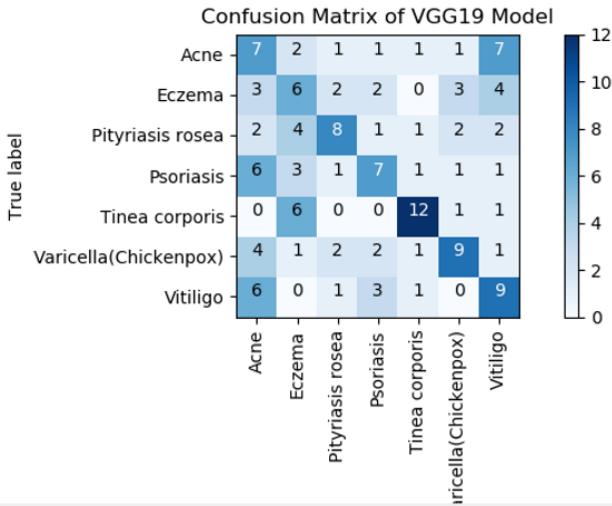

Figure 9. VGG19 Confusion Matrix

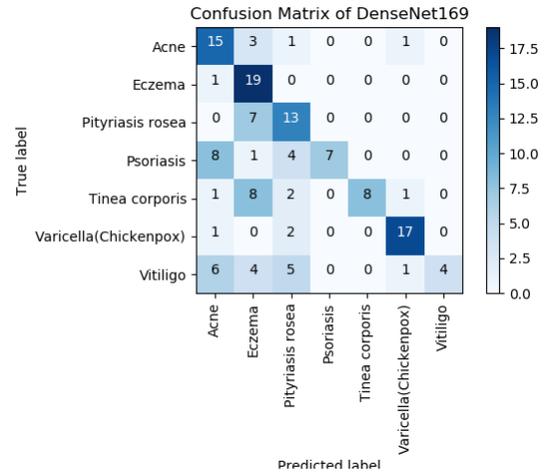

Figure 11. DenseNet169 Confusion Matrix

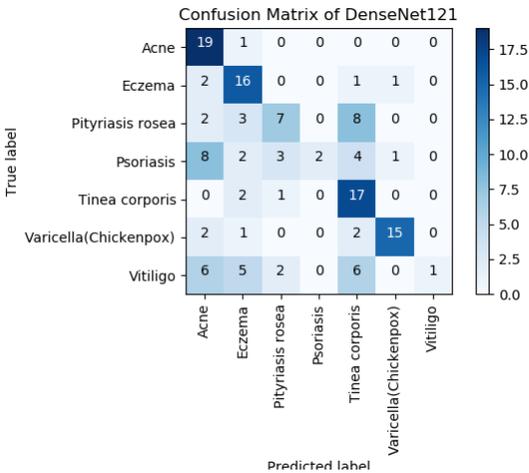

Figure 10. DenseNet121 Confusion Matrix

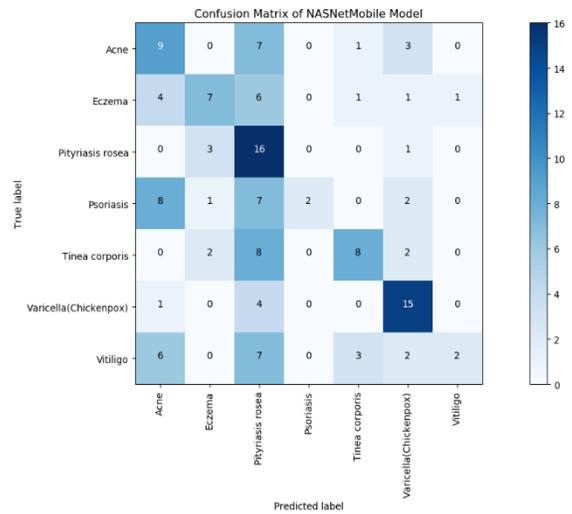

Figure 12. NASNet Mobile Confusion Matrix

*B. Weight Size, Loading time and Accuracy*

As shown in Table II, the model that performs the best is MobileNet, which has an accuracy of 94.1% and a weight size of 16.823MB, making it the model with the highest accuracy and smallest weight size. However, when compared to MobileNet, VGG16 and VGG19 load pages in 3.543 and 3.809 seconds, respectively.

TABLE II
EVALUATION OF EACH MODELS BY ITS WEIGHT SIZE, LOADING TIME AND ACCURACY

| Model | Weight size | Loading time (seconds) | Accuracy (Percent) |
|---|---|---|---|
| MobileNet | 16.823MB | 4.838 | 94.1 |
| VGG16 | 540.496MB | 3.543 | 41.4 |
| VGG19 | 561.242MB | 3.809 | 41.42 |
| Xception | 89.731MB | 6.381 | 54.29 |
| ResNet50 | 100.443MB | 7.126 | 61.4 |
| InceptionV3 | 93.860MB | 11.896 | 48.6 |
| InceptionResNetV2 | 219.932MB | 57.227 | 52.86 |
| DenseNet121 | 33.317MB | 20.787 | 64.3 |
| DenseNet169 | 58.441MB | 47.153 | 59.3 |
| DenseNet201 | 82.110MB | 10.533 | 49.29 |
| NASNet Mobile | 23.660MB | 7.657 | 42.14 |

V. CONCLUSION

The MobileNet model outperforms the others with an accuracy of 94.1% and a weight size of 16.823MB. It offers the highest precision and the smallest weight size. VGG16 and VGG19, on the other hand, load pages faster than MobileNet, taking 3.543 and 3.809 seconds, respectively.

*Acknowledgement*

The authors would like to express their acknowledgement for the assistance of the following: Jean Wilmar Alberio, Jonathan Apuang, John Stephen Cruz, Mark Angelo Gomez, Benjamin Molina Jr., and Lyndon Tuala, for their utmost contribution and effort on the completion of this study.

REFERENCES

[1] Dayan, G.H., Quinlisk, M.P., Parker, A.A., Barskey, A.E., Harris, M.L., Schwartz, J.M.H., Hunt, K., Finley, C.G., Leschinsky, D.P., O'Keefe, A.L. and Clayton, J., 2008. Recent resurgence of mumps in the United States. New England Journal of Medicine, 358(15), pp.1580-1589.

[2] Jain, P., Pathak, N., Tapashetti, P. and Umesh, A.S., 2013, December. Privacy preserving processing of data decision tree based on sample selection and singular value decomposition. In 2013 9th international conference on information assurance and security (IAS) (pp. 91-95). IEEE.

[3] Andreu-Perez, J., Poon, C.C., Merrifield, R.D., Wong, S.T. and Yang, G.Z., 2015. Big data for health. IEEE Journal of Biomedical and Health Informatics, 19(4), pp.1193-1208.

[4] Velasco, J., Pascion, C., Alberio, J.W., Apuang, J., Cruz, J.S., Gomez, M.A., Molina Jr, B., Tuala, L., Thio-ac, A. and Jorda Jr, R., 2019. A smartphone-based skin disease classification using Mobilenet CNN. arXiv preprint arXiv:1911.07929.

[5] Torrey, L., Walker, T., Shavlik, J. and Maclin, R., 2005. Using advice to transfer knowledge acquired in one reinforcement learning task to another. In Machine Learning: ECML 2005: 16th European Conference on Machine Learning, Porto, Portugal, October 3-7, 2005. Proceedings 16 (pp. 412-424). Springer Berlin Heidelberg.

[6] Rodrigues, C.A.D.S., Vinhal, C. and da Cruz, G., 2017, November. Fully convolutional networks for segmenting images from an embedded camera. In 2017 IEEE Latin American Conference on Computational Intelligence (LA-CCI) (pp. 1-6). IEEE.

[7] Xia, X., Xu, C. and Nan, B., 2017, June. Inception-v3 for flower classification. In 2017 2nd international conference on image, vision and computing (ICIVC) (pp. 783-787). IEEE.

[8] Kim, W., Choi, H.K., Jang, B.T. and Lim, J., 2017, October. Driver distraction detection using single convolutional neural network. In 2017 international conference on information and communication technology convergence (ICTC) (pp. 1203-1205). IEEE.

[9] Zhang, X., Li, Z., Change Loy, C. and Lin, D., 2017. Polynet: A pursuit of structural diversity in very deep networks. In Proceedings of the IEEE conference on computer vision and pattern recognition (pp. 718-726).

[10] Kumar, A.S. and Sherly, E., 2017, April. A convolutional neural network for visual object recognition in marine sector. In 2017 2nd International Conference for Convergence in Technology (I2CT) (pp. 304-307). IEEE.

[11] Ali, M., Sahin, F., Kumar, S. and Savur, C., 2017, October. 360° view camera based visual assistive technology for contextual scene information. In 2017 IEEE International Conference on Systems, Man, and Cybernetics (SMC) (pp. 2135-2140). IEEE.

[12] G. Huang, Z. Liu, L. van der Maaten, and K. Weinberger, ''Densely connected convolutional networks,'' in Proc. CVPR, Honolulu, HI, USA, Jul. 2017, pp. 2261–2269

[13] Stehman, S.V., 1997. Selecting and interpreting measures of thematic classification accuracy. Remote sensing of Environment, 62(1), pp.77-89.